\RequirePackage{fix-cm}
\documentclass[smallextended]{svjour3}       % onecolumn (second format)
\smartqed  % flush right qed marks, e.g. at end of proof
\usepackage{graphicx}
\usepackage{mathptmx}      % use Times fonts if available on your TeX system
\usepackage[numbers,super,comma,sort&compress]{natbib}
\usepackage{ulem,color}
\journalname{Journal of Low Temperature Physics}
\newcommand{\onlinecite}[1]{\hspace{-1 ex} \nocite{#1}\citenum{#1}} 
\begin{document}

\title{Prediction for two spatially modulated superfluids: $^4$He on fluorographene
and on hexagonal BN%\thanks{Grants or other notes
}


\author{Pier Luigi Silvestrelli \and
        Marco Nava \and
	Francesco Ancilotto \and
	Luciano Reatto
}


\institute{Pier Luigi Silvestrelli \at
              Dipartimento di Fisica e Astronomia ``Galileo Galilei'' and CNISM, Universit\`a di Padova, via Marzolo 8, 35122 Padova, Italy and 
	      CNR-IOM Democritos, via Bonomea, 265 - 34136 Trieste, Italy
	      \and
           Marco Nava \at
             Department of Chemistry, University of Illinois, Urbana, Illinois 61801, United States 
	     \and
	     Francesco Ancilotto \at
	     Dipartimento di Fisica e Astronomia ``Galileo Galilei'' and CNISM, Universit\`a di Padova, via Marzolo 8, 35122 Padova, Italy and 
	      CNR-IOM Democritos, via Bonomea, 265 - 34136 Trieste, Italy
	     \and
	     Luciano Reatto \at
          Dipartimento di Fisica, Universit\`a degli Studi di Milano, via Celoria 16, 20133 Milano, Italy
}

\date{Received: date / Accepted: date}

\maketitle

\begin{abstract}
We have derived the adsorption potential of $^4$He atoms on fluorographene (GF), on graphane and on 
hexagonal boron nitride (hBN) by a recently developed ab initio method that incorporates the van der Waals interaction. 
The $^4$He monolayer on GF and on hBN is studied by state-of-the-art quantum simulations at $T=0$ K. With our adsorption 
potentials we find that in both cases the ground state of $^4$He monolayer is a fluid and not an ordered state with localized atoms as 
on graphite and on graphene. In the case of GF the present result is in qualitative agreement with the superfluid phase that was 
obtained using an empirical adsorption potential [M. Nava et al., Phys. Rev. B 86, 174509 (2012)]. 
This fluid state of $^4$He on GF and on hBN is characterized by a very large density modulation and at the equilibrium 
density the ratio $\Gamma$ between the largest and the smallest local density along the direction of two neighboring 
adsorption sites and averaged over the perpendicular direction is $\Gamma$ = 1.91 for GF and $\Gamma$ = 1.65 for hBN. 
Recent experiments [J. Nyeki et al., Nature Physics 13, 455 (2017)] have discovered a superfluid phase in the 
second layer $^4$He. This is a spatially modulated superfluid that turns out to have anomalous thermal properties. 
This gives a strong motivation for an experimental study of monolayer $^4$He on GF and on hBN that we predict 
to be a superfluid with a much stronger spatial modulation.

\end{abstract}

\section{Introduction}
\label{intro}

%Your text comes here. Separate text sections with
%\section{Section title}
%\label{sec:1}
%Text with citations \cite{RefB} and \cite{RefJ}.
%\subsection{Subsection title}
%\label{sec:2}
%as required. Don't forget to give each section
%and subsection a unique label (see Sect.~\ref{sec:1}).
%\paragraph{Paragraph headings} Use paragraph headings as needed.
%\begin{equation}
%\end{equation}

Graphene fluoride (GF)\cite{Nai10},
also called fluorographene, and graphane (GH)\cite{Sof07},
that have been recently
obtained experimentally as chemical derivatives of graphene, 
are promising materials
for applications in many fields, but also 
represent testbed
substrates for investigating adsorption properties
of gases and liquids.
Because GF and GH have surface
symmetries and compositions which are quite different from
bare graphene (abbreviated Gr) and graphite, adsorbed gases will have very different properties
on such substrates.
Another layered material that has attracted large experimental interest is hexagonal Boron Nitride (hBN). 
hBN is an insulating isomorph of graphene, formed by alternating Boron and Nitrogen
atoms. Also in this case we can expect properties of the adsorption potential of atoms on hBN to 
be quite different from those on graphene. 

The properties of a large variety of
adsorbates are known in the case of interaction
with Gr, but not much is known
about atomic/molecular adsorption on GF, GH or hBN.
Of great interest is the behavior on such substrates of light atomic adsorbates, like
H$_2$, $^4$He and $^3$He, that is controlled by quantum fluctuations effects, and
may undergo Bose-Einstein condensation, and thus
display superfluid behavior. This expectation has been supported by a recent theoretical study\cite{Nav12} 
of $^4$He adsorbed on GF and on GH based on an empirical adsorption potential\cite{Nav12b}. 

The physics of quantum films have been largely based so far 
on detailed experimental and theoretical studies of 
the properties of He and H$_2$ on graphite substrate, 
and more recently, on theoretical studies for the graphene substrate\cite{Das94}.
In this case the ground 
state of the He film on graphite is a 2D crystal commensurate 
with the substrate, with a $\sqrt{3}\times \sqrt{3}R30^\circ$ structure.
This ordered phase (at density $\rho = 0.0636\,\AA ^{-2}$),
corresponds to atoms localized
on second nearest-neighbor hollow sites located above hexagons of C atoms.
This solid-like phase 
is known to be non-condensate (i.e. non superfluid)\cite{Buz02}.
The phase 
structures of $^4$He and para-H$_2$ films (predicted by Quantum Monte Carlo methods)
on one side and both sides of graphene\cite{Kwo12} have been shown
to be similar to that on graphite\cite{Abr87}.
A completely different behavior has been found\cite{Nav12} for a monolayer of $^4$He atoms 
adsorbed on GF and GH: the $\sqrt{3}\times \sqrt{3}R30^\circ$  ordered 
state\footnote{In the present paper the order of the adsorbed atoms is referred to the periodicity of the sheet of C atoms} 
turns out to be unstable toward a fluid state and the ground state of a monolayer of $^4$He atoms 
was found to be a spatially modulated superfluid. That result is understood in the following way: 
The adsorption sites on GF and GH are twice as many as those on GF or graphite and the energy landscape 
of He on GF and GH substrates is characterized by a very large corrugation with narrow channels along which low potential
barriers are present. Localization of $^4$He in an adsorption site cost a large kinetic 
energy so that the He atoms become delocalized and visit
only these channels, as though the atoms move in a multiconnected space along the
bonds of a honeycomb lattice. As a result, an unprecedented strongly spatially anisotropic superfluid
phase should appear, whose properties are markedly different from those of an
ordinary quasi-bidimensional quantum fluid\cite{Nav12,Nav12b,Nav12c}.
Such a novel phase has not been predicted or observed previously on any substrate
(more details on the behavior of monolayer quantum gases on graphene, graphane and
fluorographene can be found in the recent review, and references therein, by Reatto
et al\cite{Rea13}.).
The remarkable predictions made in Ref. \onlinecite{Nav12} and Ref. \onlinecite{Nav12c}
are based on accurate Quantum Monte Carlo simulations.
However, a basic ingredient of such simulations, i.e. 
the He-substrate interaction potential,
was modeled using a traditional
semiempirical approach\cite{Sto80}, where the potential energy of a
single He atom near the surface is written
as a sum of pair potential interactions 
with different layers of the substrate
for the attractive
part, and a repulsive part proportional to the local
electron density.
These empirical potentials are known to 
be affected by 
quite large uncertainties in the empirical coefficients
used to model the interaction.
For this reason, we decided to investigate from first principles
the interaction of He atoms with graphane (GH) and fluorographene (GF), 
using state-of-the-art functionals
specifically designed to describe the weak VdW interactions,
with the goal of providing a more accurate description
of the interaction of He atoms with these surfaces.
We also studied the interaction of He with a
monolayer of hexagonal BN (hBN).

Numerous methods 
have been developed in recent years
to include VdW interactions within density-functional theory (DFT),
with the goal of modeling 
van der Waals interactions
for atoms and molecules on surfaces.
(for a comprehensive review on the subject, see Ref.
\onlinecite{Ber15}, Ref. \onlinecite{Klimes} and references therein).

Recent applications of vdW-corrected DFT schemes to the
problem of atoms/molecules-surface interactions
have proven the accuracy of such methods
in the calculation of both adsorption distances and adsorption 
energies, as well as the high degree of its reliability 
across a wide range of adsorbates.

In sect. \ref{sec2} we derive ab-initio the adsorption potential of He on GF, GH and hBN. 
In sect. \ref{sec3} we describe our many-body computations of $^4$He on GF and hBN and show 
that the $^4$He atoms are delocalized on these substrate so that the system does not form an ordered 
state but it is a highly anisotropic fluid. Our conclusions are in sect. \ref{sec4}.

\section{The adsorption potential}\label{sec2}

All calculations have been performed
with the Quantum-ESPRESSO ab initio package\cite{ESPRESSO}.
A single He atom is considered and we model the substrates
adopting periodically repeated orthorhombic supercells,
with a $4 \times 2$ structure, in the case of GF and GH, 
of 32 carbon atoms plus as many F or H atoms.In the case of hBN the substrate 
is formed by 16 Boron and 16 Nitrogen atoms. The
lattice constants correspond to the equilibrium state of the substrates. 
Repeated slabs were separated along the direction orthogonal to the surface by a vacuum region
of about 24 \AA\ to avoid significant spurious interactions due to periodic
replicas. The Brillouin Zone has been sampled using a
$2\times2\times1$ $k$-point mesh.
Electron-ion interactions were described using ultrasoft
pseudopotentials and the wavefunctions were expanded in a plane-wave basis 
set with an energy cutoff of 51 Ry.

The calculations have been performed by adopting different 
DFT functionals: the PBE
Generalized Gradient Approximation (GGA) functional\cite{PBE},
which nowadays probably represents the most popular DFT functional,
the DFT-D2\cite{Grimme} functional, where vdW
corrections are implemented at a semiempirical level, and the 
rVV10\cite{Sabatini} functional (this is the revised, more efficient 
version of the original VV10 scheme\cite{Vydrov}), where
vdW effects are included by introducing an explicitly nonlocal correlation 
functional. 
rVV10 has been found to perform well in many systems and processes 
where vdW effects are relevant, including several adsorption 
processes\cite{Sabatini,psil15,psil16}.
All the tested DFT functionals are able to well reproduce the reference
structural data of graphene, graphane and fluorographene and hBN, including
the ``buckling displacement'' in GH and GF\cite{Nav12}.

\begin{figure*}
  \includegraphics[width=0.75\textwidth]{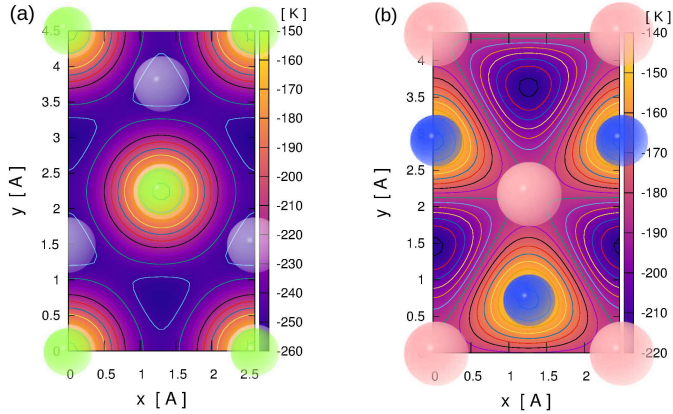}
\caption{Minimum potential energy map for (a) GF and (b) hBN. A top-view of 
the atomic lattice of the substrate as been superimposed with the following coloring for the atoms. For 
GF: (yellow) Fluorine, (gray) Carbon. For hBN: (pink) Boron, (blue) Nitrogen.}
\label{fig1}       
\end{figure*}

We have calculated the equilibrium positions and adsorption 
energies of He on various high symmetry sites on the GF, GH and hBN substrates
such as the {\it hollow} site,
on the center of a triplet of F(H) atoms for the case of
GF(GH) (or at the center of a B-N ring in the case of
hBN), the {\it top} sites,
on top of a C, H, F, B or N atom, and {\it bridge} sites between 
high symmetry sites. A representation of the minimum potential energy surface 
of He for both GF and hBN is shown in Fig. \ref{fig1}
\begin{table}
\caption{Binding energy in the lowest-energy configuration, $E_b$, 
distance $d$ of He from the reference plane (defined by averaging over
the $z$ coordinates of the C atoms for GF and GH and of the B and N for hBN), 
maximum corrugation, $\Delta_{\rm max}$, minimum
intersite energy-barrier, $\Delta_{\rm min}$ (see text for the definitions),
for He-Gr, He-GH, He-GF and for He-hBN. For He-Gr the data of Ref. \cite{Nav12} actually
refer to He on graphite.} 
\begin{tabular}{|c|l|r|c|r|r|}
\hline\noalign{\smallskip}
 system & method          & $E_b$(K) & $d$ &$\Delta_{\rm max}$(K) & $\Delta_{\rm min}$(K) \\ 
\hline
 He-Gr  & PBE             &    -77   & 3.24&      17   &       16   \\ 
   "    & DFT-D2          &   -307   & 2.95&      56   &       52   \\ 
   "    & rVV10           &   -298   & 2.96&      50   &       47   \\ 
   "    &Ref. \onlinecite{Nav12}&   -224   & 2.6 &      43   &       41   \\ 
\hline
 He-GH  & PBE             &    -88   & 4.21&      15   &        1   \\ 
   "    & DFT-D2          &   -209   & 3.77&      40   &        7   \\ 
   "    & rVV10           &   -196   & 3.81&      27   &        6   \\ 
   "    &Ref. \onlinecite{Nav12}&   -195   & 3.7 &      50   &       13   \\ 
\hline
 He-GF  & PBE             &    -95   & 4.34&      15   &        3   \\ 
   "    & DFT-D2          &   -287   & 4.07&      56   &       12   \\ 
   "    & rVV10           &   -259   & 4.02&      51   &       11   \\ 
   "    &Ref. \onlinecite{Nav12}&   -496   & 3.6 &     130   &       24   \\ 
\hline
 He-hBN & DFT-D2          &   -300   & 2.95&      52   &       25   \\ 
 
   "    & rVV10           &    -219   & 2.96 &  36     &  17      \\ 
 
\noalign{\smallskip}\hline
\end{tabular}
\label{table-energy}
\end{table}
Our numerical results are summarized in Table \ref{table-energy}.
For comparison, we also show the results for the bare graphene substrate.
We report the distance $d$ of He from the substrate and the binding 
energy $E_b$ of the lowest-energy configurations,
which is the {\it hollow} site for He on graphene; the same is true for 
He on GF and GH, where, however this configuration is essentially
isoenergetic with that corresponding to He on {\it top} of a C atom
(the difference in energy is smaller than 2 K), in line with the
results of Ref. \onlinecite{Nav12}. As a result the relevant adsorption sites are twice as many as those 
occurring on Gr and graphite. 
The importance of properly describing vdW effects is evident: in fact,
in the case of He on graphene, where reliable reference data are available,
the PBE functional, which does not reproduce vdW interactions,
dramatically underestimate $E_b$ and overestimate $d$.  
Among the tested vdW-corrected DFT functional, rVV10 turned out to give
the better performances (for instance, the semiempirical DFT-D2 approach
tends to overbinding). 
To better characterize the adsorption of He on the different substrates,
in Table I we also report two other energetic parameters: the 
``maximum corrugation'', $\Delta_{\rm max}$, defined as the difference between 
the binding energy of He on top of C, H, and F (which represents the 
less-favored configuration for He-Gr, He-GH, and He-GF, respectively) and the
binding energy of the lowest-energy configuration, and the 
``minimum intersite energy-barrier'', $\Delta_{\rm min}$, 
which is given by the minimum 
energy-barrier that the He atom must overcome to be displaced from
an optimal adsorption site to another, namely from hollow to hollow
for He-Gr, and from hollow to top-C for He-GH and He-GF. This latter
quantity has been evaluated by monitoring the binding-energy 
corresponding to a reaction path generated by constraining the 
planar $x,y$ coordinates of the He atom and optimizing the 
$z$ vertical coordinate only.  
In the case of hBN, $\Delta_{\rm max}$ ($\Delta_{\rm min}$)
correspond to the difference between
the binding energy of He on top of the N(B) atom and the 
binding energy of the lowest-energy configuration, respectively.

As can be seen, the most striking difference between the case of He-Gr
and those of He-GH and He-GF, is that in He-Gr $\Delta_{\rm max}$ 
and $\Delta_{\rm min}$ are
comparable, while, on the contrary, in He-GH and He-GF $\Delta_{\rm min}$ is 
much smaller than $\Delta_{\rm max}$.
This result is in qualitative agreement with the findings of Nava
{\it et al.}\cite{Nav12}, although our predicted $\Delta_{\rm min}$ 
values are quantitatively even smaller than those predicted in Ref. \onlinecite{Nav12}. 
This confirms that the adsorption potential of He on GH and GF is characterized by narrow ``canyons'' 
between adsorption sites, with a much larger anisotropy in the corrugation and a relatively low energy barrier 
compared to Gr and graphite. A large difference between $\Delta_{min}$ and $\Delta_{max}$ is found also for 
hBN so that the corrugation is larger than in the case of Gr. In addition around an adsorption site the saddle 
points are 3 and not 6 as in the case of Gr. 

Besides the lowest-energy configurations for 
a given investigated adsorption site,
we have also computed the dependence upon the 
normal coordinate $z$ of the He-substrate interaction potentials 
above those sites.
Our goal is to provide a 
reliable three-dimensional potential function $V_{He-s}({\bf r})$
which could be used for simulations of the behavior 
of He films on such substrates, like the ones 
possible, e.g., using the phenomenological DFT approach to the
properties of inhomogeneous $^4$He systems\cite{Anc17} 
or Quantum Monte Carlo simulations as we do in the present paper.
We approximate the potential $V_{He}(r)$ by using a truncated Fourier expansion
over the first three stars of the two-dimensional reciprocal lattice associated 
with a triangular lattice with a two-atom basis (one C and one F atom in the case 
of GF, one B and one N atom in the case of hBN). 
The Fourier components can be easily obtained from the calculated z-dependence of the various symmetry sites described above.

\section{Monolayer of $^4$He on GF and hBN}\label{sec3}
The 3-D potential $V_{He-s}({\bf r})$ has been used in 
Quantum Monte Carlo simulations based on the ground state path integral method\cite{sarsa}, 
an unbiased $T=0$ K method for bosons\cite{patatelesse} with the goal of providing
evidence of possible superfluid behavior of monolayer
$^4$He adsorbed on the studied substrates. The first step is to verify that a superfluid phase 
is not preempted by an ordered state with localized atoms as is the case of graphite and Gr. 
Therefore this initial exploration focuses on the stability of the $\sqrt{3}\times \sqrt{3}R30^\circ$ phase
that could exist at density $\rho = 0.0574$~\AA$^{-2}$ for GF and $\rho = 0.0606$~\AA$^{-2}$ for hBN
(the different densities arise from the slight difference of the lattice parameters of GF, hBN and graphite). 
We find that at such densities $^4$He on GF and on hBN form 
a self-bound state with binding energies, respectively, $E_{b-GF}= 1.0(1)$K and 
$E_{b-hBN}=0.91(9)$K per atom, the former should be compared with the value obtained in 
Ref. \onlinecite{Nav12} with an empirical adsorption potential, E$_{b-GF}^{se}=1.49(6)$K.

\begin{figure*}
  \includegraphics[width=0.75\textwidth]{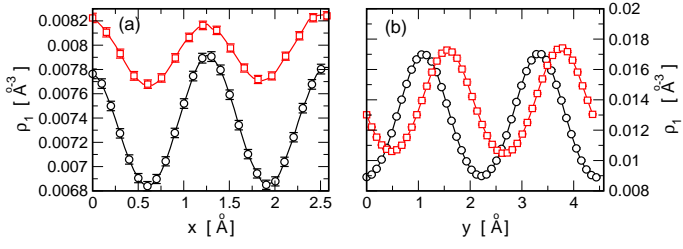}
\caption{Local density projected on (a) the $x$ direction and (b) the $y$ direction as 
depicted in Fig. \ref{fig1}. Black circles are for $^4$He on GF and red squares for $^4$He on hBN.}
\label{figdens}       
\end{figure*}

We have studied the structural properties at these special densities, $\rho=0.0574$~\AA$^{-2}$ for GF 
and $\rho=0.0606$~\AA$^{-2}$ for hBN. In both cases  we find 
a very structured density profile with modulations 
in the local density (Fig. \ref{figdens}) that have the same periodicity of the corrugation 
of the underlying substrate. These modulations can be quantified with the ratio $\Gamma$ between 
the density at a peak and that at a through of the local density, for GF this value is $\Gamma_{GF}=1.91(1)$ and 
for hBN $\Gamma_{hBN}=1.65(1)$.

\begin{figure*}
  \includegraphics[width=0.75\textwidth]{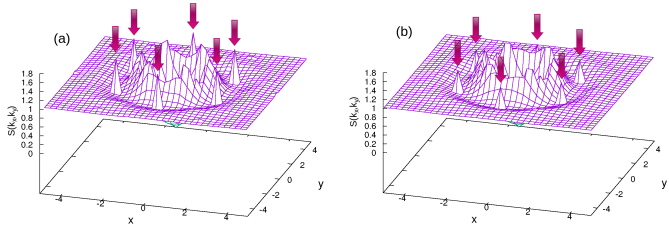}
\caption{Static structure factor for wave-vectors relative to the adsorption surface for $N=30$ atoms of $^4$He on GF (a) and 
hBN (b). The arrows indicate the peaks corresponding to the density modulation given by the substrate.}
\label{fig_sdk}       
\end{figure*}

\begin{figure*}[h!]
  \includegraphics[width=0.75\textwidth]{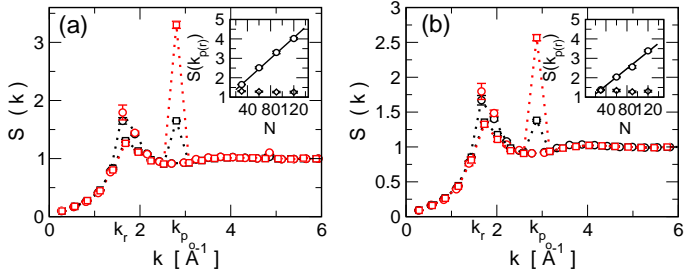}
\caption{Static structure factor for $^4$He on (a) GF and (b) hBN at different system sizes: Circles refer to $k_x$ direction, Squares 
to $k_y$ (as seen in Fig. \ref{fig_sdk}). Black color is for $N=30$ and red color for $N=90$. In the insets the intensity of the 
substrate modulation peaks (circles) and that of the ridge (diamonds) is plotted against the particle number $N$, the solid lines are linear fits to the data.}
\label{fig_scaling}       
\end{figure*}

If an ordered $\sqrt{3}\times \sqrt{3}R30^\circ$ phase were stable the local density should have a repetition 
period different from that of the substrate and this is at variance from the result in Fig. \ref{figdens}. 
The absence of the $\sqrt{3}\times \sqrt{3}R30^\circ$ order can be more clearly 
seen from the static structure factor, S(k). In fact the Bragg peaks with the smallest 
$\left|\bf{k}\right| \sim 1.7$~\AA$^{-1}$ corresponding to a $\sqrt{3}\times \sqrt{3}R30^\circ$ order are 
absent in Fig. \ref{fig_sdk} and in their place we find a ridge in k space. Such ridge denotes short range order. 
The only Bragg peaks present in S(k) are those corresponding  to the periodicity of the substrate.  
Notice the different scaling with the number $N$ of particles of a Bragg peak (intensity proportional to $N$) and 
of a peak due to short range order (intensity independent on $N$). Our results for the peak intensities, 
shown in Fig. \ref{fig_scaling} for different values of $N$ clearly 
display such different scaling. The present results for GF are in qualitative 
agreement with those obtained in Ref. \onlinecite{Nav12}

We conclude that for both substrates at low coverage of $^4$He the $\sqrt{3}\times \sqrt{3}R30^\circ$ ordered state 
is unstable and the $^4$He is in a fluid state characterized by a very strong spatial anisotropy. 
Therefore we predict the existence of two new superfluids and the explicit computation of the superfluid 
fraction and of the amount of BEC is a topic of further study.

\section{Conclusions}\label{sec4}

We have used advanced DFT theories with vdW corrections to determine the adsorption potential for 
He on three substrates: GF, GH, and hBN. In all three cases the adsorption potential differs in a 
significant way from that for graphite and graphene due to a larger corrugation and to a strong 
spatial anisotropy around the adsorption sites. Our many-body quantum simulations show that the 
most outstanding effect is that the ordered $\sqrt{3}\times \sqrt{3}R30^\circ$ state of $^4$He 
is unstable on these substrates and at low coverage $^4$He is in a fluid state characterized by 
a strong spatial modulation due to the substrate potential. Therefore we predict the existence 
of new spatially anisotropic superfluids. In the case of GF the present results are in qualitative, 
but not in quantitative, agreement with the results of Ref. \onlinecite{Nav12} based on an empirical adsorption potential. 

The prediction of a spatially anisotropic superfluid is especially interesting in view of the recent results\cite{nyeki} 
 on the behavior of the second layer of $^4$He on graphite. In this case the modulation of $^4$He in the second layer 
is mainly determined by the crystalline order of the $^4$He atoms of the first layer. A superfluid phase has been 
detected experimentally in the second layer with an exotic phase transition to the normal phase. Our predicted superfluid phase 
of $^4$He monolayer on GF and hBN has a spatial modulation much larger than that expected for the second layer on graphite 
so it should be of great interest an experimental verification of our prediction.

We are not aware of experimental study of the phase diagram of $^4$He adsorbed on hBN  while some studies 
have been performed for 3He\cite{3he1,3he2,3he3}. From these NMR measurements the authors conclude that at low temperature $^3$He 
forms an ordered state and this is identified as the $\sqrt{3}\times \sqrt{3}R30^\circ$  phase. Our study is for $^4$He, not $^3$He, 
but at first sight those experiments seem in contradiction with our results because it is difficult to understand how $^3$He could have a 
stable $\sqrt{3}\times \sqrt{3}R30^\circ$ phase while we find that for the heavier mass $^4$He such phase is unstable. 
A first comment is that the mentioned NMR measurements\cite{3he1,3he2} give evidence that at a special coverage the $^3$He 
atoms are in registry with those of hBN but there is no direct evidence that this ordered phase of $^3$He corresponds indeed to a 
coverage of 1/3 of the adsorption sites as in the $\sqrt{3}\times \sqrt{3}R30^\circ$  phase. On the basis of adsorption isotherms 
this NMR special coverage can be reconciled\cite{3he4} with the $\sqrt{3}\times \sqrt{3}R30^\circ$  phase only by advocating 
the presence of ill controlled strongly binding sites and edge effects. Also the presence of a liquid component\cite{3he2}, 
in addition to the solid one, over a large range of temperature and coverage is difficult to understand 
if the $\sqrt{3}\times \sqrt{3}R30^\circ$  phase were to lowest energy state of the monolayer as it is in 
the case of graphite. Our results show that the $\sqrt{3}\times \sqrt{3}R30^\circ$  phase of $^4$He on a monolayer of 
hBN is unstable, this does not exclude that a different commensurate ordered phase might be stable at a coverage larger than 1/3. 
A second comment is that these measurements\cite{3he1,3he2,3he3} are performed on powder of hBN and it is known\cite{3he4} 
that the adsorption properties have a significant dependence on the preparation method of the powder. In any case the single platelets 
in such powders have thickness of a fraction of micron so they are formed by many layers of the basal plane. 
Our study is for a single layer of hBN. Our computation of the adsorption potential can be extended to the case of a multilayer hBN 
and it will be interesting to verify if the instability of the $\sqrt{3}\times \sqrt{3}R30^\circ$  phase of $^4$He remains true also 
for the multilayer. In conclusion, further experimental and theoretical work seems warranted on the adsorption of the He isotopes on hBN.

\section{Acknowledgements}\label{ack}
We thank John Saunders for bringing Ref. \onlinecite{3he1} to our attention. 

This is the original paper submitted to JLTP. The published paper can be found at 
https://doi.org/10.1007/s10909-018-02133-y

%\begin{figure}
%  \includegraphics{example.eps}
%\end{figure}
%\begin{figure*}
%  \includegraphics[width=0.75\textwidth]{example.eps}
%\end{figure*}
%\begin{table}
%\begin{tabular}{lll}
%\hline\noalign{\smallskip}
%\end{tabular}
%\end{table}

%\begin{acknowledgements}
%\end{acknowledgements}

\bibliographystyle{spphys}       % APS-like style for physics


\end{document}